\documentclass[]{spie}  %>>> use for US letter paper
\usepackage[utf8]{inputenc}
\usepackage[frenchb]{babel}
\usepackage{newunicodechar}
\usepackage{textcomp}
\usepackage{siunitx}

\usepackage[frenchb]{babel}
\usepackage{amssymb, amsmath}
\usepackage[pdftex]{graphicx}
\usepackage{bmpsize}
\DeclareGraphicsExtensions{.jpg} 
\DeclareTextSymbol{\degre}{T1}{6}
\DeclareTextSymbol{\degre}{OT1}{23}
\newunicodechar{°}{\degree}

\title{A near-infrared SETI experiment: probability distribution of false coincidences}

\author{J\'er\^ome Maire\supit{a}, Shelley A. Wright\supit{a,b}, Dan Werthimer\supit{c}, Richard R. Treffers\supit{d},\\ Geoffrey W. Marcy\supit{e}, Remington P. S. Stone\supit{f}, Frank Drake\supit{g}, Andrew Siemion\supit{e}
 \skiplinehalf
\supit{a}Dunlap Institute for Astronomy \& Astrophysics, University of Toronto, 50 St. George Street, Toronto, ON M5S 3H4, Canada; \\
\supit{b}Astronomy and Astrophysics Dept, University of Toronto, 50
St. George Street, Toronto, ON M5S 3H4, Canada;\\
\supit{c} Space Sciences Laboratory, University of California Berkeley, Berkeley, CA 94720, USA\\
\supit{d} Starman Systems, LLC, Alamo, CA 94507, USA\\
\supit{e} Astronomy Department,  University of California Berkeley, Berkeley, CA 94720, USA\\
\supit{f} Lick Observatory, University of California Santa Cruz, Santa Cruz, CA 95064 USA \\
\supit{g} SETI Institute, Mountain View, CA 94043 USA }
\authorinfo{Further author information: (Send correspondence to J.M. and/or S.A.W.)\\J.M.: E-mail: maire @ dunlap.utoronto.ca\\  S.A.W.: E-mail: saw @ dunlap.utoronto.ca}
\begin{document}
  \maketitle

%%%%%%%%%%%%%%%%%%%%%%%%%%%%%%%%%%%%%%%%%%%%%%%%%%%%%%%%%%%%%
\begin{abstract}
A Search for Extraterrestrial Life (SETI),
based on the possibility of interstellar communication via laser
signals, is being designed to extend the search into the
near-infrared spectral region (Wright et al, this conference). The dedicated near-infrared (900 to
1700 nm) instrument  takes advantage
of a new generation of avalanche photodiodes (APD), based on
internal discrete amplification. These discrete APD (DAPD) detectors have a high
speed response ($>$ 1 GHz) and gain comparable to photomultiplier
tubes, while also achieving significantly lower noise than previous
APDs. We are investigating the use of DAPD detectors in this new
astronomical instrument for a SETI search and transient source
observations. We investigated experimentally the advantages of
using a multiple detector device operating in parallel to remove
spurious signals. We present the detector characterization and
performance of the instrument in terms of false positive detection
rates both theoretically and empirically through lab measurements.
We discuss the required criteria that will be needed for laser light
pulse detection in our experiment. These criteria are defined to
optimize the trade between high detection efficiency and low false
positive coincident signals, which can be produced by detector dark
noise, background light, cosmic rays, and astronomical sources. We
investigate experimentally how false coincidence rates depend on
the number of detectors in parallel, and on the signal pulse height
and width. We also look into the corresponding threshold to each of the signals to optimize the sensitivity while also reducing the false coincidence rates. Lastly, we discuss the analytical solution used to predict the probability of laser pulse detection with multiple detectors.

%short abstract:
%A Search for Extraterrestrial Life (SETI) instrumentation program,
%based on the possibility of interstellar communication via laser
%signals, is being designed to extend the search into the
%near-infrared spectral region. The dedicated Near-InfraRed and
%Optical SETI (NIROSETI) instrument (Wright et al, this conference)
%takes advantage of a new generation of avalanche photodiodes (APD),
%based on internal discrete amplification. These APD detectors have a
%high speed response (> 1 GHz) and gain comparable to photomultiplier
%tubes, while also have significantly lower noise than previous APDs.
%We present the detector characterization and performance of the
%instrument in terms of false positive detection rates both
%theoretically and as tested directly through lab measurements.

\end{abstract}

%>>>> Include a list of keywords after the abstract


%%%%%%%%%%%%%%%%%%%%%%%%%%%%%%%%%%%%%%%%%%%%%%%%%%%%%%%%%%%%%
\section{INTRODUCTION}
\label{sec:intro}  % \label{} allows reference to this section

Scientific searches for signs of
extraterrestrial intelligence have been undertaken since the mid-twentieth century. Discoveries of exoplanets starting in the mid-1990's have further strengthened the possibility of life elsewhere in the galaxy (Mayor \& Queloz 1995 \cite{Mayor95}). Several stars have been identified that host planets that are the right size and distance from their stars to support liquid water, and thus life as we know it (Petigura et al. 2013\cite{Petigura13}, Quintana  et al. 2014\cite{Quintana14}, Anglada-Escud\'e et al. 2014\cite{Anglada14}).
Programs, such as SETI, are conducting an astronomical search for
radio activity which would confirm the presence of intelligent life.
It is also possible that a distant civilization might send out pulsed and
continuous laser signals in the optical, as well as infrared,
spectrum (Schwartz \& Townes 1961 \cite{Schwartz1961}). Dan Werthimer's group at UC Berkeley tested the first pulsed
multiple detector system in 1997 in the optical domain, bringing forth OSETI programs (Werthimer et al. \cite{Werthimer2001}, Howard et al. 2000\cite{Howard2000}, Horowitz et al. 2001\cite{Horowitz2001}, Wright et al. 2004\cite{Wright2004}, Stone et al. 2005\cite{Stone2005}, Drake et al. 2010\cite{Drake2010}). 

  The Near-InfraRed Optical Search for Extraterrestrial Intelligence (NIROSETI), (Wright et al.
  \cite{Wright2014}, this conf.), is a dedicated near-infrared instrument, currently under development, that will be used primarily to search for
near-infrared (0.95$\mu$m to 1.65$\mu$m) nanosecond pulsed laser signals, that could have been
transmitted by a pulsed laser operated by a distant civilization. In addition to this program, the NIROSETI instrument will also be used to study variability of very short natural near-infrared transients over a temporal range stretching from seconds to nanoseconds.

  Astrophysical sources of nanosecond flashes of significant intensity appear to be unknown.
When detecting light pulses from the neighborhood of a star,
different sources of noise can occur, including the light from the
star itself. The use of a fast  detector at the nanosecond
timescale, gives the ability to detect single photon arrival, and
considerably decrease the probability of having several star photon
detections in a nanosecond window.  The light collected by the
telescope can be separated via beam splitters to be detected by
several detectors working in coincidence, giving the capability to
suppress single-detector false events. The detection of a nanosecond
light pulse can be defined as a sufficient number of
photo-events  in a nanosecond window. The choice of this
threshold is crucial to determine false alarms and probability of
pulse detection. We will discuss the probability of false alarm rate
in Section \ref{sec:1} using lab measurements.
In Section \ref{sec:2}, we study the probability of detection of a pulsed laser signal. Then, we will infer the probability of detection of a laser pulse with constant false alarm rate in Section \ref{sec:3}.
%%%%%%%%%%%%%%%%%%%%%%%%%%%%%%%%%%%%%%%%%%%%%%%%%%%%%%%%%%%%%

\section{False Alarm rate}\label{sec:1}
One important aspect of nanosecond laser pulse detection is the
ability to distinguish the signal of interest against  background
noise.  In the case of the optical and NIR SETI instruments, different sources
of noise, such as pulses generated by the incoming star light, sky
background and detector intrinsic noise can yield  false alarms.
Other potential sources of noise, such as Cerenkov radiation due to
cosmic-ray muons, or direct ejection of an electron from the
photo-cathode due to the muon traveling through all detectors haven't
shown any evidence of  events for optical SETI \cite{Howard2004}.

The way the NIROSETI detectors work is that if two photons are detected within the integration time 
(approximately 0.5ns), then the output voltage of the pulse will be
doubled. It is possible to define a voltage threshold above which any pulse or coincidence can be considered to probably originate from a target. The voltage threshold should be sufficiently high to suppress false alarm events.

Since false events occurring on different detectors are independent,
it is possible to calculate the chances of coincident false events
by multiplying the chances of false events happening on each
detector. If the light is split equivalently onto the detectors
and if the detectors produce an equivalent number of dark pulses per
second, the coincident false alarm rate can be defined
as (Wright et al. 2001\cite{Wright2001})
 \begin{equation}
    \label{eq1}
R = \left(\frac{r}{N} + d \right)^N  t_{c}^{N-1}
   \end{equation}
where N is the number of detectors, r is the number of photon-events
per second due to star and sky-background light, d is the number of
dark pulses per second generated by each detector,  and $t_c$ is the
time during which such event are detected.  This will be referred
below as the Wright et al. model. In the following subsections, we explore the false alarm rate through lab
measurements.

%%-----------------------------------------------------------
\subsection{Pulse height distribution} \label{subsec:dark}

The number of dark pulses per second, d in Eq.1, should be chosen as
small as possible to limit false alarms. The NIROSETI instrument is
equipped with near-infrared Discrete Amplification Photo-Diode
(DAPD\cite{Linga2005, rafi2008}), which are InGaAs photo-detectors designed for
the analog detection of extremely low-level light signals (from one
photon to several thousand photons), in the (950nm to 1650nm) spectral
range.
Photo-electrons generated in the photo-conversion process are
 divided into N equal charge components. Each
individual elementary charge is directed into an individual channel
of a multichannel threshold amplifier. Due to discretization of the
input signal into equal charge components and subsequent calibration
of a charge packet in each channel of the threshold amplifier, it is
possible to attain an extremely low noise factor of amplification \cite{Linga2005, rafi2008}.
The value of the Excess Noise Factor (ENF), referring to noise due to
the amplification process, could be close to unity even at
very high gain.  
Use of discrete amplification technology allows proportional
detection of a light signal with  high gain ($>$1e5), fast
response ($<$ 0.5ns rise time) and negligible excess noise factor ($<$
1.05).
 %-------------
  \begin{figure}
  \begin{center}
  \begin{tabular}{c}
  \includegraphics[height=5cm,natwidth=452,natheight=210]{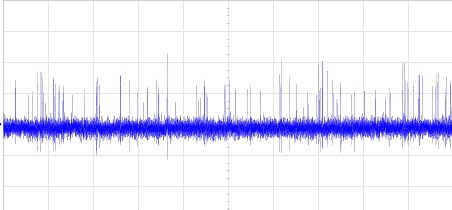}
  \includegraphics[height=5cm,natwidth=560,natheight=420]{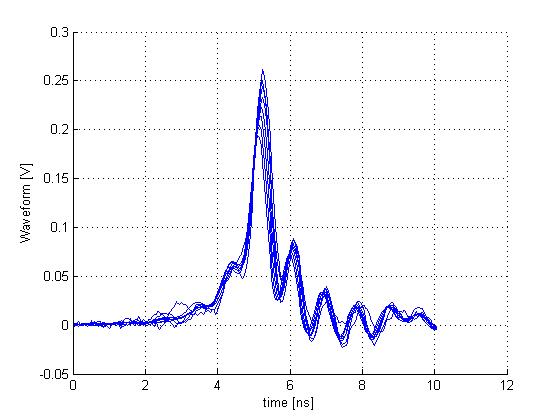}
  \end{tabular}
  \end{center}
  \caption{Left: typical $80\mu m$-DAPD long waveform (100ns per division) showing dark pulses. Right: on a shorter timescale, typical mean dark extracted from a one millisecond acquisition time. }
%>>>> use \label inside caption to get Fig. number with \ref{}
  \label{fig:pulses}
  
  \end{figure}
%-------------

This DAPD device is biased  with a constant
voltage, where the bias voltage determines the
sensitivity\cite{Linga2005}.  The output of the DAPD detector is
connected through a capacitor, thus the DC element of the current is
not flowing into the sensing electronic amplifier, only the pulse
response, which is a response either to photon absorption or a dark pulse.
This device is quite special in that the dark pulse is the same as
the single electron pulse (SER).  SER is the response for absorbing
one photon. For DAPD, ENF represents basically the variation in gain of the
SER. 
The operating conditions are determined by the field inside the
avalanche region, which in turn is determined by the DC voltage
applied on the device.  This device is designed to operate in the
so-called Geiger mode, where the avalanche is in breakdown
condition, in an operating voltage that is above the breakdown
voltage.  The breakdown conditions are actually controlled in this
device, and the current that flows above breakdown is regulated.
This is in contrast to a conventional APD,  which will theoretically
reach infinite current above breakdown if kept in a DC bias above
the breakdown bias.  However, the DAPD has a mechanism that limits
this current.

\begin{figure}
  \begin{center}
  \begin{tabular}{c}
  \includegraphics[width=12cm,natwidth=560,natheight=420]{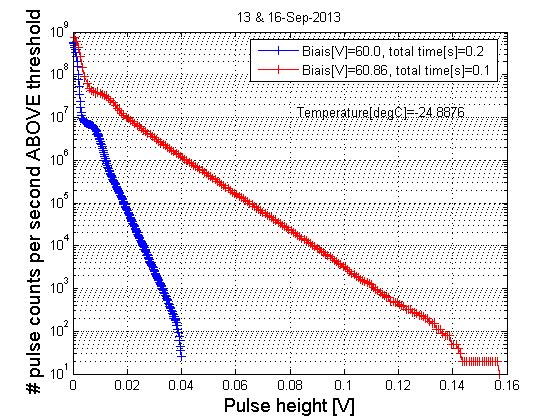} \\
  \includegraphics[width=12cm,natwidth=1024,natheight=587]{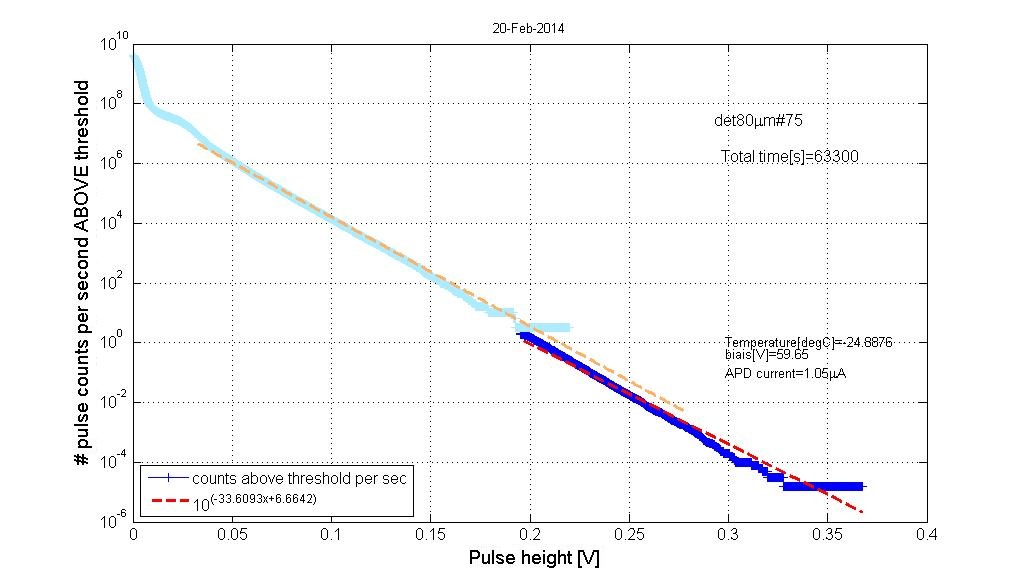}
  \end{tabular} %width=\textwidth
  \end{center}
  \caption{Top: Number of dark pulses per second having height above a threshold given by the voltage on x-axis, for two different biases applied to the DAPD. Bottom: Number of dark pulses per second above threshold, measured by analysis of the waveform (in cyan), and by analysis of scope trigger events (in blue). Dashed lines represents lines of best fit, according to the model described in Section \ref{subsec:coinc}}
%>>>> use \label inside caption to get Fig. number with \ref{}
 \label{fig:dist}
 
  \end{figure}

The number of dark pulses per second generated by the DAPD detector
have been measured in the lab using an RF amplifier and a 2.5 GHz
scope to sample the waveform at a rate of 20Gsps, and a high-voltage
regulated power supply. Detectors have been cooled at
-25\si{\degree}C by use of
temperature controllers. 
Figure \ref{fig:pulses} (left) illustrates a
typical DAPD waveform over one microsecond period. All dark pulses are
saved and can be used to produce a mean dark pulse in some specific height
ranges (Fig.\ref{fig:pulses}, right) that would be proportional to the SER. 
Pulses can be detected using two different practical approaches: by post-analysis of the entire saved waveform or by triggering pulse events using the scope. Analysis of the waveform gives the capability to detect every single peak automatically by finding a change of sign in the first derivative of the saved signal.  Once the peak is detected, its output voltage, or pulse height can be measured. We can then determine a threshold, and count all the
pulses that cross this threshold over a period of time.
Figure \ref{fig:dist} (Left) represents the number of dark pulses per second having height above a threshold given by the voltage on x-axis, for two different biases applied to the DAPD. The knee of the curve at small pulse heights is due to scope intrinsic noise fluctuations increasing the number of very small pulse numbers. Recording waveforms at a 20Gsps sampling rate requires large amounts of memory and disk space. Triggering of pulse events directly from the scope is a more convenient method  for detecting pulses over a longer period of time. Only pulses higher than the trigger threshold are detected, with a maximal rate of 3 per second, due to hold-off and saving time occurring after the trigger event. The triggering method allowed us to analyse dark pulse height over 20 to 30 hours period of time and to confirm that the slope of the distribution in log scale stays constant over long period of time. Only one outlier has been reported during one of the two 20-30 hours long run measurements, that could possibly have been caused by a cosmic ray.

 %-------------
%-------------

\subsection{Pulse width}

\begin{figure}
  \begin{center}
  \begin{tabular}{c}
  \includegraphics[width=10cm,natwidth=1067,natheight=833]{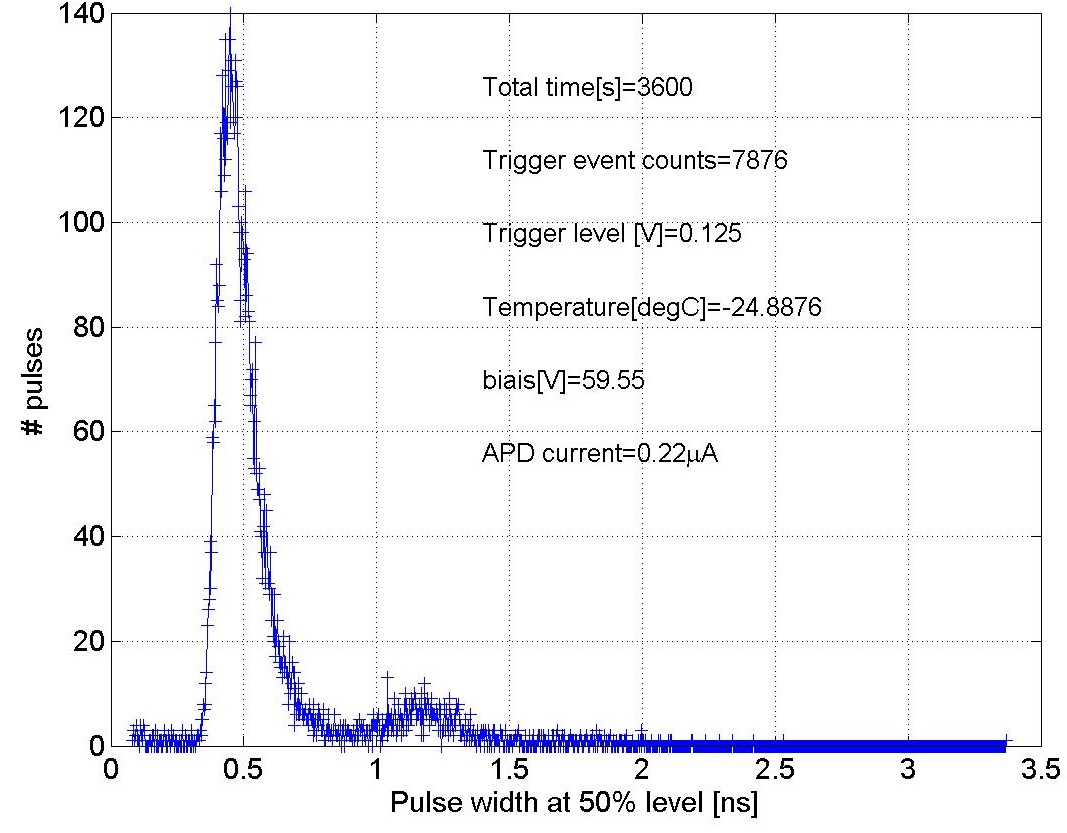}
  \end{tabular}%\textwidth
  \end{center}
  \caption {Distribution of pulse widths ( 80$\mu$m DAPD), measured with dark pulses that triggered the scope (for peak voltages $>$ 0.125V) during one hour of measurements. Most of the pulses have about 0.5ns FWHM. }
%>>>> use \label inside caption to get Fig. number with \ref{}
  \label{fig:width}
 
  \end{figure}

The amplification process of this DAPD device causes dark pulses to appear identical to the single electron pulse.  The excess noise factor  gives fluctuations of the pulse shape and width. We measured widths of individual pulses with high height using the scope trigger to detect dark pulses over a period of one hour. Figure \ref{fig:width} represents the measured distribution of pulse width obtained using an $80\mu m$-DAPD. 
Most of the pulses have 0.5ns FWHM. This value defines the coincidence time, if  pulses occur within this interval on each  detector.

%-------------
%-------------

\subsection{Dark pulse coincidence} \label{subsec:coinc}
\begin{figure}
  \begin{center}
  \begin{tabular}{c}
  \includegraphics[width=\textwidth,natwidth=997,natheight=542]{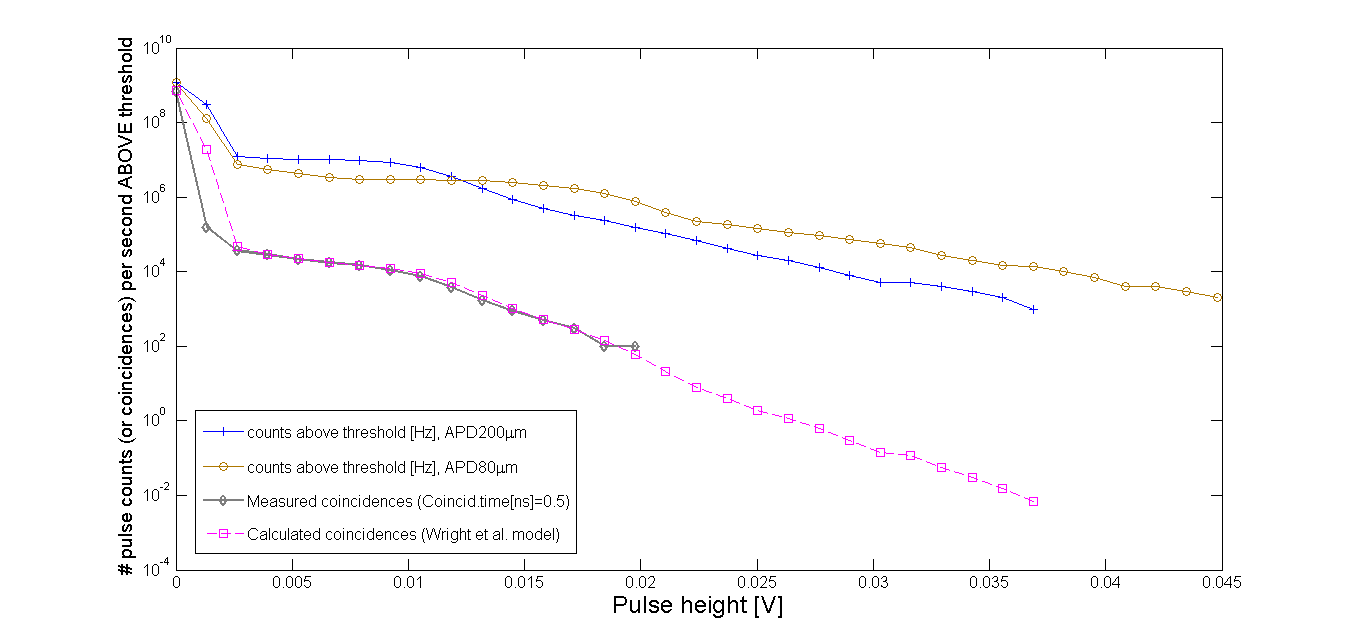}
  \end{tabular}
  \end{center}
  \caption{Measured number of dark pulse coincidences between the two channels (diamond markers). Individual dark pulse height  distributions of each DAPD are represented with crosses and circle markers.The deduced false alarm rates given by the Wright et al. model (square) fit the measured coincidences.  }
%>>>> use \label inside caption to get Fig. number with \ref{}
 \label{fig:coinc}
 
  \end{figure}

We used two DAPDs simultaneously  to measure coincident dark pulses and test the Wright et al. model. Since the sampling time is shorter than the coincident time,  output waveforms are first smoothed using a time window that has the width of a pulse (0.5ns). Pulses are then detected on the two smoothed waveforms. There is a coincidence if the locations of the pulse maxima in the two waveforms fall within the coincidence time. Figure \ref{fig:coinc} represents the measured dark pulse height distributions of the two DAPDs ($d_1$ and $d_2$),  the measured number of coincidences between the two channels and the coincidence function given by the Wright et al. model defined in this case as $d_1\times d_2\times t_{c}$. As the figure illustrates, a very good match were found, the model giving an excellent estimate of the measured false alarm rate. This has been tested for different detector biases, and each time confirmed the relevance of the Wright et al. model.

It is also possible to obtain an analytical expression of the dark pulse height distribution tail for a given DAPD bias. By fitting a function of the form $d(V)=10^{aV+b}$ to the tail of the distribution, one can obtain the slopes $a_1$, $a_2$ and intercepts $b_1$, $b_2$ for each of the two detectors.

 %-------------TO DO: ADD figure 2 waveforms with coincidence
%-------------

%%%%%%%%%%%%%%%%%%%%%%%%%%%%%%%%%%%%%%%%%%%%%%%%%%%%%%%%%%%%%
%%%%%%%%%%%%%%%%%%%%%%%%%%%%%%%%%%%%%%%%%%%%%%%%%%%%%%%%%%%%%
\section{Probability of light pulse detection}\label{sec:2}
As seen in the previous section, it is possible to decrease the false alarm rate (FAR) by increasing the pulse height  threshold or by working in coincidence logic with several detectors, but
 at the expense of decreasing the probability of detection of a laser light pulse.  When one works in coincidence logic with several detectors, the beam of light is split into two or several channels, decreasing the number of photons hitting each individual detector. Even if the beam splitter is made to have 50\% transmitted light and 50\% reflected light, it is not guaranteed that each light pulse will have its number of photons exactly divided by two on each channel, especially when the total number of photons is low. We consider hereafter the probability of light pulse coincidence when  randomly assigning a small set of photons into a group of detectors. 

If n is the number of photons arriving, within a maximal time equal to the coincidence time $t_c$, on the system containing k detectors, there is a coincident detection of this light pulse if the detectors are all receiving at least 1 photon (within $t_c$).   We are interested in the probability P that every detector receives at least 1 photon (within $t_c$
). In other words, if we write X for the number of "empty" detectors receiving no photon, the probability of coincidence is P(X=0). The solution to this occupancy problem is using the Stirling number of the second kind S(n,k), or
the number of ways to partition the set of n photons into k
non-empty detectors. We can partition the photons $\left\{1,2,\ldots,n\right\}$ into k non-empty sets, then assign the detectors $\left\{1,2,\ldots,k\right\}$ to these sets. %\ldots
There are S(n,k) partitions, and for each partition, there are $k!$ ways to assign the detectors. Furthermore, there are $k^n$ equally likely ways to distribute the n photons among the k detectors. Thus,
 \begin{equation}
    \label{eq2}
P(X=0) = \frac{S(n,k)k!}{k^n}
   \end{equation}
or equivalently
 \begin{equation}
    \label{eq3}
P(X=0) = \frac{\left(\frac{1}{k!}\sum\limits_{j=0}^{k}(-1)^{k-j}C^n_kj^n\right)k!}{k^n}
   \end{equation}
   where $C^n_k$ is the binomial coefficient ($C^{n}_{k} = \frac{n!}{k!(n-k)!}$). Equation \ref{eq3}
can be simplified as 
 \begin{equation}
    \label{eq4}
P(X=0) = \frac{\sum\limits_{j=0}^{k}(-1)^{k-j}C^n_kj^n}{k^n}
   \end{equation}
The numerator of Eq.\ref{eq4} counts all possible
ways to have detection, whereas the denominator counts all the ways
to distribute the n photons among the k detectors.

\begin{itemize}

  \item Case 2 detectors: Putting k=2 in Eq.\ref{eq4} gives
 \begin{equation}
    \label{eq5}
P(X=0) = \frac{-C^2_1+C^2_2 2^n}{2^n}
   \end{equation}
Thus, for 2 detectors we have:
 \begin{equation}
    \label{eq6}
P(X=0) = \frac{ 2^n - 2}{2^n}
   \end{equation}
  \item Case 3 detectors: Putting k=3 in Eq.\ref{eq4} gives
 \begin{equation}
    \label{eq7}
P(X=0) = \frac{C^{3}_{1}-C^3_2 2^n+C^3_3 3^n}{3^n}
   \end{equation}
Thus for 3 detectors, we have
 \begin{equation}
    \label{eq8}
P(X=0) = \frac{3-3\times 2^n+ 3^n}{3^n}
   \end{equation}
  \item Case 4 detectors: Putting k=4 in Eq.\ref{eq4} gives
 \begin{equation}
    \label{eq9}
P(X=0) = \frac{-C^4_1+C^4_2 2^n-C^4_3 3^n+C^4_4 4^n}{4^n}
   \end{equation}
Thus for 4 detectors, we have
 \begin{equation}
    \label{eq10}
P(X=0) = \frac{-4+6\times 2^n-4\times 3^n+ 4^n}{4^n}
   \end{equation}
\end{itemize}

\begin{figure}
  \begin{center}
  \begin{tabular}{c}
  \includegraphics[width=12cm,natwidth=446,natheight=322]{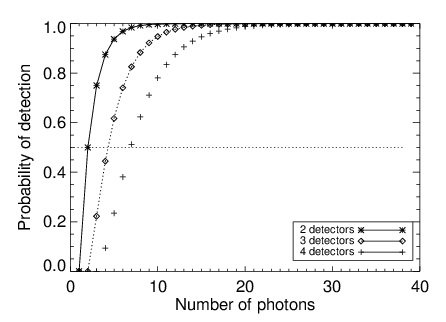}
  \end{tabular}%\textwidth
  \end{center}
  \caption {Probability that every detector receives at least 1 photon (no thresholding) as a function of the number of photons in the pulse. The dashed line represents 50\% probability of coincidence.}
%>>>> use \label inside caption to get Fig. number with \ref{}
  \label{fig:prob1}
 
  \end{figure}

These probabilities of coincidence are represented in Figure \ref{fig:prob1}. The 50\% probability of coincident detection is reported in Table \ref{tab:proba}. For a given number of photons in a pulse, the probability of coincident detection decreases with the number of detectors.

%-------------
%-------------

 We have defined above a coincident detection of
light pulse where the detectors are all seeing at least 1 photon (within $t_c$). We could also define a detection with thresholding if every detector get at least r photons. What is the probability $P_r$
that every detector get at least r photons (within $t_c$)? In other
words, if we write X for the number of empty detectors, what is the
probability $P_r(X=0)$ that each detector receives at least r photons?
We will use an r-associated Stirling number of the second kind (Howard 1980\cite{Howard1980}),
or the number of ways to partition a set of n photons into k detectors, with each detector receiving at least r photons. Using the same arguments than above, we can show that the probability $P_r$ that
every detector get at least (r+1) photons is given by
 \begin{equation}
    \label{eq11}
P_r(X=0) = \frac{S_r(n,k)k!}{k^n}
   \end{equation}
or equivalently 
\begin{equation}
    \label{eq12}
P_r(X=0) = \frac{\left( (-1)^k n!\sum\limits_{k_1,k_2,\dots,k_{r+2}} \frac{(-1)^{k_1} (k_1)^A}{k_1!k_2!\ldots k_{r+2}!A!Q} \right)k!}{k^n}
   \end{equation}

where the sum all over $k_1,k_2,\dots,k_{r+2}$ is such that $k_1+k_2+\dots+k_{r+2} = k$
with $0 \le k_i \le k$, and with\\
$A=A\left(k_1,k_2,\dots,k_{r+2}\right)=n-\sum\limits_{i=0}^{r}i k_{i+2}$

and \\
$Q=Q\left(k_1,k_2,\dots,k_{r+2}\right)=\prod\limits_{i=0}^{r}(i!)^{ k_{i+2}}$

It can be shown that Eq.\ref{eq4} is a particular case of Eq.\ref{eq12} with r=0.

\begin{figure}
  \begin{center}
  \begin{tabular}{cc}
  \includegraphics[width=8cm,natwidth=482,natheight=314]{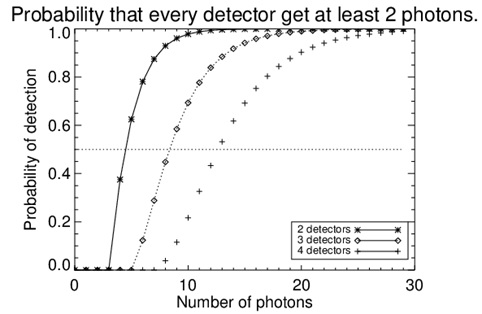} &
   \includegraphics[width=8cm,natwidth=461,natheight=303]{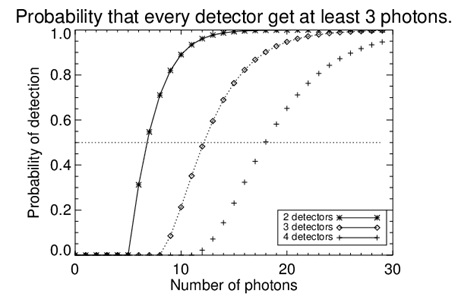} \\
   \includegraphics[width=8cm,natwidth=443,natheight=286]{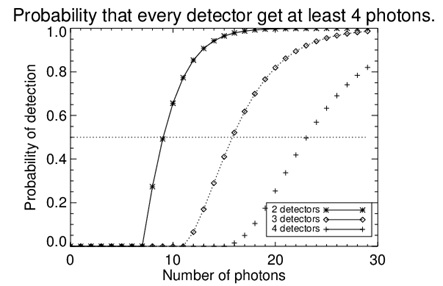} &
      \includegraphics[width=8cm,natwidth=443,natheight=298]{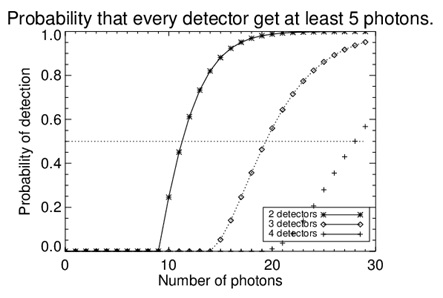}
  \end{tabular}%\textwidth
  \end{center}
  \caption {Probability of coincidence when the threshold is set to be at least 2 photons (top left),  at least 3 photons (top right), at least 4 photons (bottom left) and at least 5 photons (bottom right) for  k=2,3 and 4 detectors in the instrument.}
%>>>> use \label inside caption to get Fig. number with \ref{}
  \label{fig:prob2}
 
  \end{figure}

Using Eq.\ref{eq12} with r=1, the probability $P_1(X=0)$ that each detector
receive at least 2 photons is
\begin{equation}
    \label{eq13}
P_1(X=0) =\frac{\left( \sum\limits_{i=0}^{k}(-1)^iC^i_n  \sum\limits_{j=0}^{k-i} \frac{(-1)^j(k-i-j)^{(n-i)}}{j!(k-i-j)!} 
\right)k!}{k^n}
   \end{equation}
Figure \ref{fig:prob2} (top left) represents the probability $P_1(X=0)$ of coincidence as a function of the number of photons when the threshold is set to be at least 2 photons. We used Eq.\ref{eq12} to calculate numerically  the probability $P_2(X=0)$ of coincidence when the threshold is set to be at least 3 photons (top right panel), $P_3(X=0)$ for the threshold to be at least 4 photons (bottom left panel), and $P_4(X=0)$ for the threshold to be at least 5 photons (bottom right panel).

%-------------
%-------------
We assumed 100\% detector efficiency for convenience, otherwise the number  of photons needs to be scaled by the inverse of the efficiency value. In the case of the 20\% DAPD efficiency,  number of photons in Table \ref{tab:proba} have to be multiplied by 5.
Multiplying the appropriate factors, the  NIROSETI instrument with two detectors at the Lick 1m telescope (with 70\% transmission) is sensitive to signals of 40 optical photons (950 -1650 nm) per square meter arriving in a group within 0.5 ns when the threshold is set to 2 photon-events.

\begin{table}[h]
\caption{Total number of photons required in the pulse to reach 50\% probability of coincidence as a function of threshold and number of detectors in the instrument. 100\% detector efficiency is assumed.} 
\label{tab:proba}
\begin{center}       
\begin{tabular}{|l|l|l|l|l|l|} %% this creates two columns
%% |l|l| to left justify each column entry
%% |c|c| to center each column entry
%% use of \rule[]{}{} below opens up each row
\hline
\rule[-1ex]{0pt}{3.5ex} threshold [nb of photons]   & 1 & 2 & 3 & 4 & 5  \\
\hline
\rule[-1ex]{0pt}{3.5ex}    2 detectors system & 2 phot. & 4.5 phot. & 6.8 phot.& 9 phot.& 11.2 phot.\\
\hline
\rule[-1ex]{0pt}{3.5ex}  3 detectors system & 4.4 phot. & 8.4 phot.& 12.2 phot.& 15.9 phot.& 19.4 phot.\\
\hline
\rule[-1ex]{0pt}{3.5ex}  4 detectors system & 6.9 phot.& 12.8 phot.& 18 phot.&  23 phot.& 28 phot.\\
\hline
\end{tabular}
\end{center}
\end{table}

\section{Probability of pulse detection with constant false alarm rate} \label{sec:3}

The choice of a constant false alarm rate constrains
the voltage threshold above which any pulse or coincidence can be
considered to probably originate from a pulsed laser. If this threshold is
low, then more targets will be detected at the expense of increased
numbers of false alarms. Conversely, if the threshold is too high,
then fewer targets will be detected, but the number of false alarms
will also be low. The threshold can be set in order to
achieve a required probability of false alarm or equivalently, false
alarm rate.

 Let's say we want to calibrate the instrument so that the false alarm rate is equal or better (smaller) than $2.7 \times 10^{-7}$Hz (1 per 1000 hours), and let's consider that false alarms are produced by detector darks only, and let's assume  100\% efficiency. Using pulse height (above threshold) measurements made in the lab on Nov. 27th 2013 on a 80$\mu$m APD, we know that the number of darks per second as a function of the pulse height voltage V is $d[V]=(10^{-57.96 \times V + 7.58} )$, for a 59V bias and 0.22$\mu$A current at -24.8\si{\degree}C. We assume that the tail of the distribution follows the same slope.
If we assume that $V_1$=21mV, which is given by the knee of the pulse height distribution, is the height of a single-photon pulse, then
\begin{itemize}

  \item for one detector, we need to set the threshold to 12 photons to have R$<$ 2.7e-7 Hz, because $d[12\times V_1]  = 9$e-8 Hz
  
 \item for two detectors,  dark pulse events from different detectors are independent, then using Eq.\ref{eq1}, the false alarm rate is $R=d^2 \times t_c$ where $t_c$ is the coincidence time (0.5ns). The condition R $<$ 2.7e-7 Hz  is met if $V>5V_1$, since $(d[5\times V_1])^2\times t_c=~4$e-7 Hz.   Fig.\ref{fig:prob2} (bottom-right panel)  shows that 50\% probability of coincidence needs 11.3 photons, if a coincidence is defined as every detector receives at least 5 photons.
 
\item for three detectors, dark pulses from different detectors are independent, then $R=d^3 \times t_c^2$. The condition $R< 2.7$e-7 Hz is met if $V>3V_1$, because $(d[3\times V_1])^3\times t_c^2=1$e-7 Hz.  Fig.\ref{fig:prob2} (top-right panel)  shows that 50\% probability of coincidence needs 12.2 photons, if a coincidence is defined as every detector receives at least 3 photons.
 
 \item for four detectors, dark pulses from different detectors are independent, then $R=d^4 \times t_c^3$. The condition $R< 2.7\times 10^{-7}$Hz 
 is met if $V>2V_1$, 
 since $(d[2\times V_1])^4 \times t_c^3=4$e-8Hz. 
 Fig.\ref{fig:prob2} (top-left)  shows that 50\% probability of coincidence needs 12.8 photons, if a coincidence is defined as every detector receives at least 2 photons.
\end{itemize}

For this example with a given detector bias and constant false alarm rate (1 per 1000 hours), we have shown that the two detectors case requires less photons to reach 50\% probability of coincident detection. However, the number of photons needed to reach the same constant false alarm rate in the 1 detector case do not rely on 50\% probability. The pulse height distribution changes with the detector bias and may be different in the presence of star and sky-background light, so the results of this example can not be generalized to choose the optimal number of detectors.

\section{Future work} \label{sec:future}
We plan to measure in the lab the  pulse height distribution given by continuous and pulsed light and test the Wright et al. model in these conditions. The optimal threshold value can be determined as a function of the chosen sensitivity. We plan to measure how faint the light pulses can be to still be detected by the NIROSETI instrument in coincident observations with two detectors, or in direct mode using one detector.

%%%%%%%%%%%%%%%%%%%%%%%%%%%%%%%%%%%%%%%%%%%%%%%%%%%%
%\appendix    %>>>> this command starts appendixes
%%%%%%%%%%%%%%%%%%%%%%%%%%%%%%%%%%%%%%%%%%%%%%%%%%%%

%%%%%%%%%%%%%%%%%%%%%%%%%%%%%%%%%%%%%%%%%%%%%%%%%%%%%%%%%%%%%
\acknowledgments     %>>>> equivalent to \section*{ACKNOWLEDGMENTS}
We would like to thank the generosity of our anonymous donor for supporting this research and instrument; this individual’s interest and support catalyzed these endeavors. We would also like to extend our thanks to Rafeal Ben-Michael from Amplifcation Technologies Inc. for  aiding our testing with the discrete amplification APDs. The Dunlap Institute is funded through an endowment established by the David Dunlap family and the University of Toronto.
%%%%%%%%%%%%%%%%%%%%%%%%%%%%%%%%%%%%%%%%%%%%%%%%%%%%%%%%%%%%%
%%%%% References %%%%%
%\nocite{*}
\bibliography{bibliospie2014c}   %>>>> bibliography data in report.bib
%

%\end{thebibliography}
\bibliographystyle{spiebib}   %>>>> makes bibtex use spiebib.bst

\end{document}